# Reversible Structural Transition in Epitaxial Manganite Film

Q. Zhan[1], R. Yu[1*], L. L. He[1], D. X. Li[1], J. Li[2], S. Y. Xu[2], and C. K. Ong[2]

[1] Shenyang National Laboratory for Materials Science, Institute of Metal Research, Chinese Academy of Sciences, 72 Wenhua Road, Shenyang 110016, China

[2] Department of Physics, National University of Singapore, Singapore 119260

**Abstract**

A reversible structural transition of an epitaxial $La_{2/3}Sr_{1/3}MnO_3$ film deposited on the $LaAlO_3$ substrate has been investigated by means of *in situ* high-resolution transmission electron microscopy (HREM) and electron diffraction, combined with image and diffraction calculations. We observe that the crystallographic symmetry of the film can be lowered via electron beam irradiation, leading to a rhombohedral-monoclinic transition. This transition can be attributed to the cooperating effect of the mismatch stress and the irradiation-induced thermal-stress.

Since the colossal magnetoresistance (CMR) effect was observed in perovskite mixed valence manganites $RE_{1-x}A_xMnO_3$ (RE, rare-earth element; A, alkaline metal), the magnetic, electronic and magnetotransport properties of these materials and their relations to microstructures have attracted considerable interest [1-2]. The studies on epitaxial films were especially developed widely and profoundly because of their special properties and applications [3]. An important difference between epitaxial films and bulk materials is that the lattice mismatch between the film and the substrate imposes a strain pattern in the system which influences its magnetotransport properties [4].

A perovskite-type $RE_{1-x}A_xMnO_3$ system rarely has an ideal perovskite structure, due to the distortion resulted from the cation size mismatch and the Jahn-Teller effect, whereby a distortion of the oxygen octahedron sorrounding the Mn cation splits the energy levels of the latter, thus lowering the energy. The magnitude of the distortion depends on the balance between the Jahn-Teller effect and the concomitant lattice strain. It is realized recently that the Jahn-Teller type electron-lattice coupling is also important to the magnetoresistance properties of the manganites [5], besides the classical double exchange mechanism [6-7].

In addition to the structural distortion, perovskite manganites exhibit a variety of magnetic, orbital and charge orderings. Transitions between these phases can be induced by external perturbations, such as temperature [8], magnetic field [9], pressure [10], current-injection [11], and irradiation with photons [12], X-rays [13] and electrons [14]. These phenomena have been attracting great interest in the context of unusual phase control of the magnetic and electronic states in the magnetic oxides [2].

In this work, we show that a reversible structural transition of a manganite film can be induced by electron beam irradiation. A model based on cooperating mismatch stress and thermal stress is proposed to explain the transition.

The ceramic target with nominal composition of $La_{2/3}Sr_{1/3}MnO_3$ (LSMO) was sintered by conventional solid-state reaction. LSMO films were deposited on (001)-$LaAlO_3$ single crystal substrate at 750 °C by pulsed laser ablation system [15]. The films were kept *in situ* for 1h before cooled down slowly. *In situ* high-resolution electron microscopy (HREM) observations were performed using a JEM-2010 electron microscope. The image simulations were carried out using the multislice method.

Low magnification observations of the cross-sectional sample indicate that the interface between the LSMO film and the LAO substrate is sharp and flat. A typical cross-sectional HREM image is shown in Fig. 1. A regular contrast is observed, which mainly consists of a square array of bright spots, spaced by 2.7Å. The image fits well with the simulated one for the rhombohedral structure in the space group $R\bar{3}c$. The corresponding selected area electron diffraction pattern (SAED, the inset in the top right hand corner) confirms that the sample exhibits the expected $R\bar{3}c$ rhombohedral structure.

The film was then irradiated by focused electron beam with a constant current density (about 65 pA/cm$^2$) and HREM images were recorded. When the irradiation time exceeded about 7 seconds, the period doubling contrast was appeared in the HREM images as shown in Fig. 2a. Near the film/substrate interface the image of the film consists of a row of bright dots alternating with a row of gray dots, displaying the periodicity of 7.7Å normal to the interface. The corresponding SAED pattern is given in Fig. 2b. Additional weak reflections near the middle of the main diffraction spots of the film can be observed. Series of EDPs were obtained by tilting the sample around $[001]_c$ zone axis, as shown in Fig. 2c-d. These weak diffraction spots in the $[001]_c$ direction are present constantly on tilting thus the possibility of double diffraction can be excluded. It is noted that the additional weak reflections do not bisect exactly the main diffraction spots of the film. Their reciprocal vectors along the doubling direction are smaller than a half of those of the LSMO film. This indicates that the lattice parameter normal to the film/substrate interface has increased significantly (about 4 %) after the electron irradiation.

More interestingly, the transformed structure was unstable in the absence of irradiation. The contrast of the doubled periodicity disappeared after the electron beam had been shifted away for about 10 minutes, and the regular contrast corresponding to $R\bar{3}c$ rhombohedral structure was restored, as shown in Fig. 3. By re-irradiating the restored area with the focused electron beam, the doubled period could be obtained again. These processes indicated that the transition is reversible.

Considering the fact that the transition induced by the electron beam irradiation occurred in a short time (about 7 s), the possibility of irradiation induced chemical changes, such as long-range diffusion of atoms, oxygen vacancies and the deficiency of cations can be ruled out. It is thus inferred that the transition is a displacive one, which usually occurs in perovskite structures. For a displacive transition, the topologic relationship of atoms remains unchanged with the space group of the production phase being the subgroup of that of the parent phase. The maximal subgroups of $R\bar{3}c$ are then considered first. The structures for the subgroups of $R\bar{3}c$ are either rhombohedral or monoclinic. Noting that three $<100>_c$ directions in the rhombohedral structure are equivalent due to the three-fold axis, the new structure is certainly not a rhombohedral one since the weak spots appeared only in one of $<100>_c$ directions. Thus the $C2/c$ monoclinic maximal subgroup of $R\bar{3}c$ is the only one remained [16]. Setting $b_m$ the unique axis and the monoclinic angle $\beta$ approximate 90°, the unit cell that gives the space group *I12/a1* is chosen. Based on the EDPs shown in Figure 2b-d, the unit cell is then determined as $4a_c \times \sqrt{2}a_c \times \sqrt{2}a_c$, giving the lattice parameters $a_m \approx 1.620$ nm, $b_m \approx 0.54$ nm, $c_m \approx 0.54$ nm and $\beta \approx 90°$. The atomic coordinates are summarized in Table I.

The simulated EDPs of the monoclinic structure with *I12/a1* space group are given in Fig. 2e-g. HREM image simulations of the monoclinic structure with the atomic positions shown in Table I were also carried out along $[011]_m$ ($[100]_c$) direction, as shown in the inset in Fig. 2a. Excellent agreements between the simulated images and the experimental ones in the EDPs and HREM images are achieved. Fig. 4 shows $[011]_m$ projection of the monoclinic structure and that of the rhombohedral ($R\bar{3}c$) structure. Clearly, it is the displacements of oxygen atoms that cause the distortions of $MnO_6$ octahedron and thus lower the symmetry from rhombohedral to monoclinic. More detailed interpretation of the images would need accurate atomic positions, which cannot indeed be determined from present HREM study.

The most important result of this study is that the LSMO film on the LAO substrate underwent reversible rhombohedral ↔ monoclinic transition induced by electron beam irradiation. Since the transition was reversible, the samples could not have been damaged during the irradiation, *i.e.* neither 'knock on' of atoms nor breaking of bonding. Therefore, the most direct effect of the irradiation on the sample is the electron beam heating. Thus a possible explanation for the rhombohedral ↔ monoclinic transition is that the monoclinic structure is a high temperature phase. However, according to the phase diagram of $La_{1-x}Sr_xMnO_3$ [1], there isn't such a monoclinic phase corresponding to the composition $La_{2/3}Sr_{1/3}MnO_3$.

The more reasonable explanation for the transition is based on the mismatch stress induced by the substrate and the thermal-stress. The mechanism is schematically shown in Fig. 5. The LSMO film is under a compressive stress when it grows epitaxially on the LAO substrate since the lattice parameters of LSMO are larger than those of LAO substrate, as shown in Fig. 5a. The electron beam is then focused on the film, leading to an increase of the temperature of the irradiated area, which will expand correspondingly. As a result, the compressive stresses, especially the in-plane ones, are enhanced, as shown in Fig. 5b. Meanwhile, the constraint to the expansion normal to the film is much smaller than parallel to the film. Under this stress and constraint condition, the film will expand to outer "free space" along the film normal (Fig. 5c), giving an increased out-of-plane lattice parameter, as detected experimentally. At the same time, the in-plane compressive stresses are reduced. Furthermore, the enhanced compressive stress might break the balance constructed between the Jahn-Teller effect and the energy-raising effect associated with the elastic strain, resulting in a different distortion of oxygen octahedra with a newly constructed balance. Eventually, the monoclinic structure is formed that produces the doubled periodicity perpendicular to the film/substrate interface. On the other hand, in the absence of the focused electron beam, the temperature and the thermal-stress in the film will decrease gradually, leading to the restoration of the rhombohedral structure.

Although it is the rhombohedral structure that transformed to a monoclinic structure in the present work, it is expected that other structures, an orthorhombic one for example, could also transform to a monoclinic structure under proper stress conditions. This provides an explanation for the existence of monoclinic microdomains widely observed in orthorhombic manganites [17-18] since the anisotropic stresses exist inevitably in epitaxial film materials and usually also in polycrystalline materials.

According to the double exchange mechanism [6-7], the transfer interaction of the $e_g$ conduction electrons (holes) in the monoclinic structure is expected to be different from that in the rhombohedral one, due to a different distortion of $MnO_6$ octahedron. Therefore, it is reasonable to expect that the monoclinic domains have different resistance from the rhombohedral matrix and eventually influence the transportation properties of materials.

In summary, we show that the crystallographic symmetry of a manganite film, $La_{2/3}Sr_{1/3}MnO_3$ on $LaAlO_3$ substrate, can be decreased via electron beam irradiation, leading to a reversible rhombohedral-monoclinic transition. It is suggested that the transition results predominantly from the cooperating effect of the mismatch stress and the irradiation-induced thermal-stress. This effect manifests a strong Jahn-Teller type electron-lattice coupling of manganites as well as the sensitivity of the crystallographic structure of manganites to stresses, especially in epitaxial thin films.

The authors gratefully acknowledge the support from the National Natural Science Foundation of China on Grant No. 50071063.

* Corresponding author.

Email address: ryu@imr.ac.cn

**Figure Captions**:

FIG. 1. Typical HREM image along $[100]_c$ direction of the LSMO film on LAO substrate before the irradiation. The insets are the corresponding SAED pattern and the simulated image of LSMO with $R\bar{3}c$ space group (thickness = 16 nm, $\Delta f$ = -40 nm)

FIG. 2. (a) The $[100]_c$ HREM image of the film after an irradiation of the focused electron beam for about 7 s and (b) the corresponding SAED pattern. (c)-(d) SAED patterns along $[130]_c$ ($[0\bar{1}2]_m$) and $[120]_c$ ($[0\bar{1}3]_m$) obtained by tilting the specimen around the $[001]_c$ ($[100]_m$) zone axis. Note that additional weak spots along $[001]_c$ direction are present constantly on tilting. (e)-(g) The simulated EDPs for the monoclinic structure with *I12/a1* space group corresponding to (b)-(d), respectively. The inset in (a) is a simulated image of the monoclinic structure along $[011]_m$ ($[100]_c$) with the atomic positions shown in Tab. I. (thickness = 16 nm, $\Delta f$ = -40 nm).

FIG. 3. The HREM image recorded after the electron beam had been shifted away from the irradiated area shown in Fig. 2a. Note that the $R\bar{3}c$ rhombohedral structure was restored.

FIG. 4. $[011]_m$ ($[100]_c$) projection of the monoclinic (*I12/a1*) structure and the corresponding projection of the rhombohedral ($R\bar{3}c$) structure.

FIG. 5. Schematic diagram of the transition mechanism of the film induced by the focused electron beam irradiation. The direction and relative strength of stresses are represented schematically by arrows. (a) The stress state before the electron irradiation:

the LSMO film is under a compressive stress. (b) The compressive stress in the film is enhanced due to the thermal-stress induced by the electron beam irradiation. (c) The film expands to outer "free space" along the film normal, accompanied by the relaxation of the stress.

TABLE I. Positional parameters for the proposed monoclinic structure (space group *I12/a1*). $a_m \approx 1.620$ *nm*, $b_m \approx 0.54$ *nm*, $c_m \approx 0.54$ *nm* and $\beta \approx 90°$.

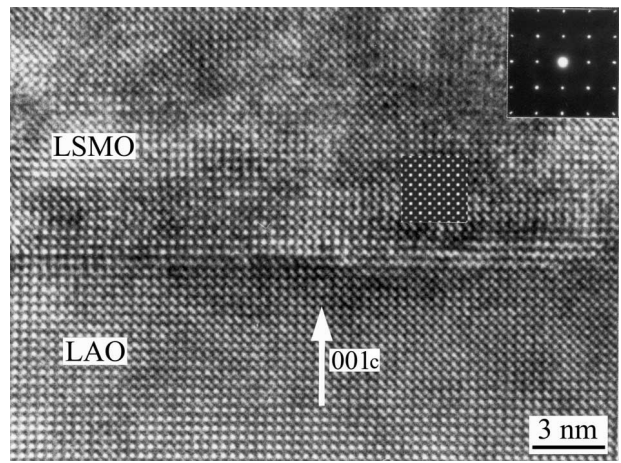

FIG. 1. Typical HREM image along [100]$_c$ direction of the LSMO film on LAO substrate before the irradiation. The insets are the corresponding SAED pattern and the simulated image of LSMO with $R\bar{3}c$ space group (thickness = 16 nm, $\Delta f$ = -40 nm)

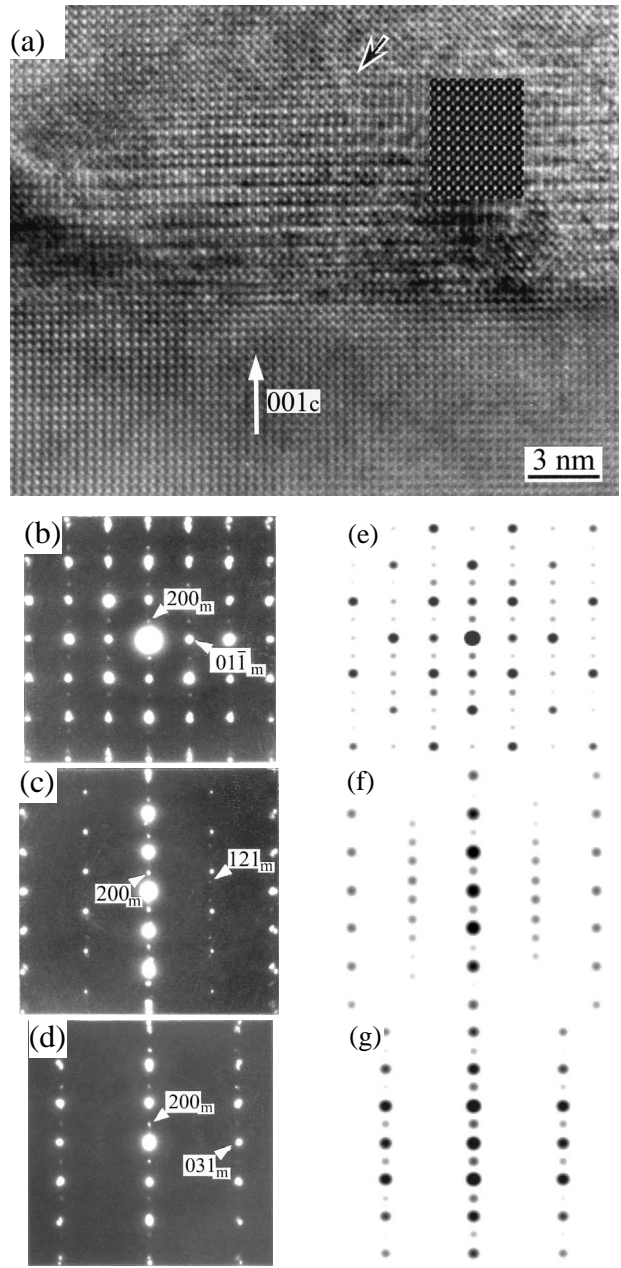

FIG. 2. (a) The $[100]_c$ HREM image of the film after an irradiation of the focused electron beam for about 7s and (b) the corresponding SAED pattern. (c)-(d) SAED patterns along $[130]_c$ ($[0\bar{1}2]_m$) and $[120]_c$ ($[0\bar{1}3]_m$) obtained by tilting the specimen around the $[001]_c$ ($[100]_m$) zone axis. Note that additional weak spots along $[001]_c$ direction were present constantly upon the tilting. (e)-(g) The simulated EDPs for the monoclinic structure with *I12/a1* space group corresponding to (b)-(d), respectively. The inset in (a) is a simulated image of the monoclinic structure along $[011]_m$ ($[100]_c$) with the atomic positions shown in Tab. I. (thickness = 16 nm, $\Delta f$ = -40 nm).

  Q.Zhan, R.Yu *et al.*

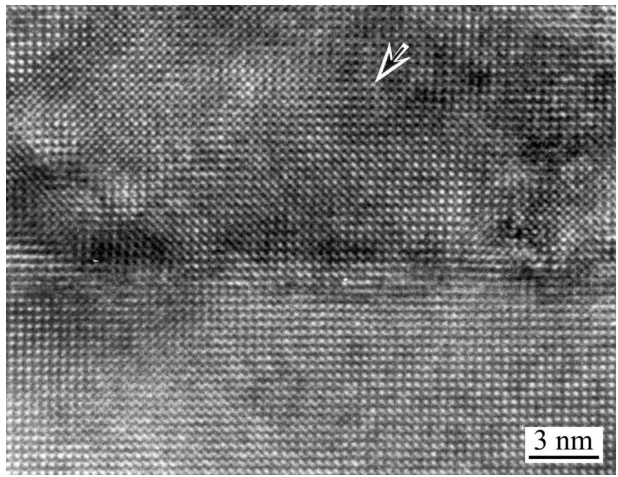

FIG. 3. The HREM image recorded after the electron beam has been shifted away from the irradiated area shown in Fig. 2a, as indicated by the arrows. Note that the $R\bar{3}c$ rhombohedral structure was restored.

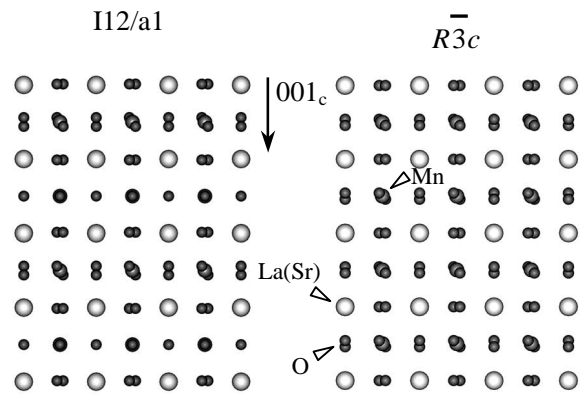

FIG. 4. $[011]_m$ ($[100]_c$) projection of the monoclinic (*I12/a1*) structure and the corresponding projection of the rhombohedral ($R\bar{3}c$) structure.

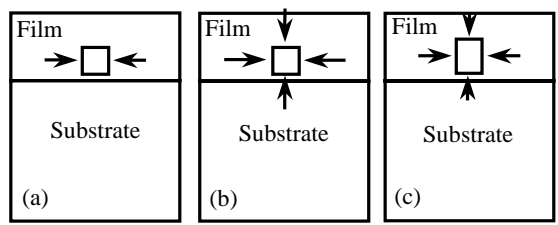

FIG. 5. Schematic diagram of the transition mechanism of the film induced by the focused electron beam irradiation. The direction and relative strength of stresses are represented schematically by arrows. (a) The stress state before the electron irradiation: the LSMO film is under a compressive stress. (b) The compressive stress in the film is enhanced due to the thermal-stress induced by the electron beam irradiation. (c) The film expands to outer "free space" along the film normal, accompanied by the relaxation of the stress.

TABLE I. Positional parameters for the proposed monoclinic structure (space group $I12/a1$). $a_m \approx 1.620$ nm, $b_m \approx 0.54$ nm, $c_m \approx 0.54$ nm and $\beta \approx 90°$.

| Atom | Wyckoff Position | x | y | z |
|---|---|---|---|---|
| Mn | 4a | 0.000 | 0.000 | 0.000 |
| Mn | 4e | 0.250 | 0.000 | 0.000 |
| La/Sr | 8f | 0.125 | 0.500 | 0.000 |
| O | 4c | 0.250 | 0.250 | 0.750 |
| O | 4d | 0.250 | 0.250 | 0.750 |
| O | 8f | 0.015 | 0.271 | 0.229 |
| O | 8f | 0.128 | 0.958 | 0.000 |